\documentclass[11pt]{article}

\newtheorem{theorem}{Theorem}
\newtheorem{lemma}[theorem]{Lemma}
\newtheorem{Corollary}[theorem]{Corollary}
\newtheorem{Claim}[theorem]{Claim}
\newcommand{\proof}{\noindent {\bf Proof: }}

\def \lket {|}
\def \rket {\rangle}
\def \lbra {\langle}
\def \rbra {|}
\newcommand{\ket}[1]{\lket #1\rket}
\newcommand{\bra}[1]{\lbra #1\rbra}
\newcommand{\qed}{\hfill{\rule{2mm}{2mm}}}
\def\A{{\cal A}}
\def\H{{\cal H}}

\def\search{SEARCH}
\newcommand{\comment}[1]{}

\begin{document}

\title{A new quantum lower bound method,
with an application to strong direct product theorem for quantum
search}
\author{Andris Ambainis\thanks{Supported by NSERC, CIAR and IQC University
Professorship.}\\
Department of Combinatorics and Optimization and \\
Institute for Quantum Computing\\
University of Waterloo\\
200 University Avenue West\\
Waterloo, ON N2L 3G1, Canada}
\date{}

\def\search{SEARCH}

\maketitle

\abstract{
We present a new method for proving lower bounds on quantum 
query algorithms. The new method is an extension of adversary
method, by analyzing the eigenspace structure of the problem.

Using the new method, we prove a strong direct product 
theorem for quantum search. This result was previously
proven by Klauck, \v Spalek and de Wolf (quant-ph/0402123)
using polynomials method. No proof using adversary method
was known before. 
}

\section{Introduction}

Many quantum algorithms (for example, Grover's 
algorithm \cite{Grover} and quantum counting \cite{Counting}) 
can be analyzed in the query model where the input is
accessed via a black box that answers queries
about the values of input bits. 

There are two main methods for proving lower bounds
on query algorithms: adversary method \cite{Ambainis00} and
polynomials method \cite{Beals+} and both of them have been
studied in detail. The limits of adversary method 
are particularly well understood. The original adversary
method \cite{Ambainis00} has been generalized in several
different ways \cite{Ambainis03,LM,BSS}. \v Spalek and Szegedy
\cite{SS} then showed that all the generalizations are 
equivalent and, for certain problems, cannot improve
the best known lower bounds. For example \cite{SS,Zhang}, 
the adversary methods of \cite{Ambainis03,LM,BSS} cannot 
prove a lower bound on a total Boolean function that exceeds
$O(\sqrt{C_0(f) C_1(f)})$, where $C_0(f)$ and $C_1(f)$ are
the certificate complexities of f on 0-inputs and 1-inputs.
This implies that the adversary methods 
of \cite{Ambainis03,LM,BSS} cannot prove a tight lower
bound for element distinctness or improve the best known 
lower bound for triangle finding. 
(The complexity of element distinctness is $\Theta(N^{2/3})$
\cite{AS,Ambainis04} but the adversary method cannot prove
a bound better than $\Omega(\sqrt{N})$.
For triangle finding \cite{MSS}, the best known lower bound is 
$\Omega(N)$ and it is known that it cannot be improved 
using the adversary method.
It is, however, possible that the bound is not tight,
because the best algorithm uses $O(N^{1.3})$
queries.)

In this paper, we describe a new version of quantum adversary
method which may not be subject to those limitations.
We then use the new method to prove a strong direct product
theorem for the {\em K-fold search} problem.

In the $K$-fold search problem, a black box 
contains $x_1, \ldots, x_N$ such that $|\{i:x_i=1\}|=K$
and we have to find all $K$ values $i:x_i=1$.
This problem can be solved with $O(\sqrt{NK})$ queries.
It can be easily shown, using any of the previously known methods, 
that $\Omega(\sqrt{NK})$ queries are required. 
A more difficult problem is to show that $\Omega(\sqrt{NK})$ 
queries are required, even if the algorithm only has to
be correct with an exponentially small probability $c^{-K}$, $c>1$.
This result is known as the {\em strong direct product theorem}
for $k$-fold search. Besides being interesting on its own,
the strong direct product theorem is useful for proving
time-space tradeoffs for quantum sorting \cite{KSW} and lower
bounds on quantum computers that use advice \cite{Aaronson}.

The strong direct product theorem for quantum search
was first shown by Klauck et al. \cite{KSW}, using polynomials
method. No proof using adversary method
has been known and, as we show in section \ref{sec:previous},
the previously known adversary methods 
are insufficient to prove a strong direct product theorem
for $K$-fold search.

\section{Preliminaries}

We consider the following problem.

{\bf Search for $K$ marked elements,} $\search_K(N)$. 
Given a black box containing $x_1, \ldots, x_N\in\{0, 1\}$ such that
$x_i=1$ for exactly $K$ values $i\in\{1, 2, \ldots, N\}$, find 
all $K$ values $i_1, \ldots, i_K$ satisfying $x_{i_j}=1$.

This problem can be viewed as computing an ${N \choose K}$-valued
function $f(x_1, \ldots, x_N)$ of variables 
$x_1, \ldots, x_N\in\{0, 1\}$, with values of the function
being indices for ${N \choose K}$ sets $S\subseteq [N]$ of 
size $K$, in some canonical ordering of those sets.

We study this problem in the quantum query model 
(for a survey on query model, see \cite{BWSurvey}).
In this model, the input bits can be accessed by queries to an oracle $X$
and the complexity of $f$ is the number of queries needed to compute $f$.
A quantum computation with $T$ queries
is just a sequence of unitary transformations
\[ U_0\rightarrow O\rightarrow U_1\rightarrow O\rightarrow\ldots
\rightarrow U_{T-1}\rightarrow O\rightarrow U_T.\]

The $U_j$'s can be arbitrary unitary transformations that do not depend
on the input bits $x_1, \ldots, x_N$. The $O$'s are query (oracle) transformations
which depend on $x_1, \ldots, x_N$.
To define $O$, we represent basis states as $|i, z\rangle$ where
$i$ consists of $\lceil \log (N+1)\rceil$ bits and
$z$ consists of all other bits. Then, $O_x$ maps
$\ket{0, z}$ to itself and
$\ket{i, z}$ to $(-1)^{x_i}\ket{i, z}$ for $i\in\{1, ..., N\}$ 
(i.e., we change phase depending on $x_i$, unless $i=0$ in which case we do
nothing). 

The computation starts with a state $|0\rangle$.
Then, we apply $U_0$, $O_x$, $\ldots$, $O_x$,
$U_T$ and measure the final state.
The result of the computation are 
$\lceil \log_2 {N \choose K} \rceil$ rightmost 
bits of the state obtained by the measurement,
which are interpreted as a description for 
one of ${N \choose K}$ subsets 
$S\subseteq \{1, \ldots, N\}$, $|S|=K$.


\section{Overview of adversary method}
\label{sec:previous}

We describe the adversary method of \cite{Ambainis00}.

Let $S$ be a subset of the set of possible inputs $\{0, 1\}^N$.
We run the algorithm on a superposition of inputs in $S$.
More formally, let $\H_A$ be the workspace of the algorithm.
We consider a bipartite system $\H=\H_A\otimes \H_I$
where $\H_I$ is an ``input subspace" spanned by
basis vectors $\ket{x}$ corresponding to inputs $x\in S$.

Let $U_T O U_{T-1} \ldots U_0$ be the sequence of unitary transformations
on $\H_A$ performed by the algorithm $A$
(with $U_0, \ldots, U_T$ being the transformations that
do not depend on the input and $O$ being the query transformations).
We transform it into a sequence of unitary transformations on $\H$.
A unitary transformation $U_i$ on $\H_A$ corresponds to
the transformation $U'_i=U_i\otimes I$ on the whole $\H$.
The query transformation $O$ corresponds to a transformation $O'$
that is equal to $O_x$ on subspace $H_A\otimes \ket{x}$.

We perform the sequence of transformations
$U'_T O' U'_{T-1}\ldots U'_0$ on the starting state
\[ \ket{\psi_{start}}=\ket{0}\otimes \sum_{x\in S} \alpha_x \ket{x} .\]
Then, the final state is
\[ \ket{\psi_{end}}= \sum_{x\in S}\alpha_x \ket{\psi_x}\otimes\ket{x} \]
where $\ket{\psi_x}$ is the final state of $A=U_T O U_{T-1} \ldots U_0$
on the input $x$. This follows from the fact that the restrictions
of $U'_T, O', U'_{T-1}, \ldots, U'_0$ to $\H_A\otimes\ket{x}$ are
$U_T$, $O_x$, $U_{T-1}$, $\ldots$, $U_0$ and these are exactly the
transformations of the algorithm $A$ on the input $x$.

Let $\rho_{end}$ be the reduced density matrix of the $\H_I$
register of the state $\ket{\psi_{end}}$. 
The adversary method of \cite{Ambainis00,Ambainis03}
works by showing the following two statements
\begin{itemize}
\item
Let $x\in S$ and $y\in S$ be such that $f(x)\neq f(y)$
(where $f$ is the function that is being computed). 
If the algorithm outputs the correct answer with
probability $1-\epsilon$ on both $x$ and $y$, then
$|\rho_{end})_{x, y}| \leq
2\sqrt{\epsilon (1-\epsilon)} |\alpha_x| |\alpha_y|$.
\item 
for any algorithm that uses $T$ queries, there are
inputs $x, y\in S$ such that $(\rho_{end})_{x, y} >
2\sqrt{\epsilon (1-\epsilon)} |\alpha_x| |\alpha_y|$
and $f(x)\neq f(y)$.
\end{itemize}
These two statements together imply that any algorithm
computing $f$ must use more than $T$ queries.

An equivalent approach \cite{HNS,Ambainis03} is to consider
the inner products $\lbra \psi_x\ket{\psi_y}$ between
the final states $\ket{\psi_x}$ and $\ket{\psi_y}$
of the algorithm on inputs $x$ and $y$.
Then, $|(\rho_{end})_{x, y}| \leq
2\sqrt{\epsilon (1-\epsilon)} |\alpha_x| |\alpha_y|$
is equivalent to $|\lbra \psi_x\ket{\psi_y}|\leq 
2\sqrt{\epsilon (1-\epsilon)}$.

As a result, both of the above statements can be described
in terms of inner products $\lbra \psi_x\ket{\psi_y}$,
without explicitly introducing the register $\H_I$.
The first statement says that, for the algorithm
to succeed on inputs $x, y$ such that $f(x) \neq f(y)$,
the states $\ket{\psi_x}$ and $\ket{\psi_y}$ must be 
sufficiently far apart one from another
(so that the inner product $|\lbra \psi_x\ket{\psi_y}|$
is at most $2\sqrt{\epsilon(1-\epsilon)}$).
The second statement says that this is impossible
if the algorithm only uses $T$ queries.

\comment{
We show that, for any algorithm that uses $T$ queries,
there are two inputs $x$, $y$, such that $f(x) \neq f(y)$
but the final states of the algorithm $\ket{\psi_x}$
and $\ket{\psi_y}$ are close. Therefore, the algorithm 
must fail to produce a correct answer on one of the inputs 
$x$ and $y$. 
}

This approach breaks down if we consider computing a 
function $f$ with success probability $p < 1/2$.
($f$ has to have more than 2 values for this task  
to be nontrivial.)
Then, $\ket{\psi_x}$ and $\ket{\psi_y}$ could be the same
and the algorithm may still succeed on both inputs,
if it outputs $x$ with probability 1/2 and $y$ with
probability 1/2.
In the case of strong direct product theorems,
the situation is even more difficult. 
Since the algorithm only has to be correct with 
a probability $c^{-K}$, the algorithm could have 
almost the same final state on $c^{K}$ different
inputs and still succeed on every one of them.

In this paper, we present a new method that does not 
suffer from this problem. Our method, described
in the next section, uses the idea
of augmenting the algorithm with an input register 
$\H_I$, together with two new ingredients:
\begin{enumerate}
\item
{\bf Symmetrization.}
We symmetrize the algorithm by applying a random
permutation $\pi\in S_N$ to the input $x_1, \ldots, x_N$.
\item
{\bf Eigenspace analysis.}
We study the eigenspaces of $\rho_{start}$, $\rho_{end}$ 
and density matrices describing the state of $\H_I$ at
intermediate steps and use them to bound the progress
of the algorithm. 
\end{enumerate}
The eigenspace analysis is the main new technique.
Symmetrization is necessary to simplify the structure
of the eigenspaces, to make the eigenspace analysis possible.

\section{Our result}

\begin{theorem}
There exist $\epsilon$ and $c$ satisfying $\epsilon>0$, $0<c<1$ 
such that, for any $K\leq N/2$, solving $\search_K(N)$ with probability 
at least $c^K$ requires $(\epsilon-o(1)) \sqrt{NK}$ queries.
\end{theorem}

\proof
Let $\A$ be an algorithm for $\search_K(N)$ that uses $T\leq \epsilon\sqrt{NK}$
queries. 

We first ``symmetrize" $\A$ by adding an extra register $\H_P$ 
holding a permutation $\pi\in S_N$. Initially, $\H_P$ holds a uniform superposition
of all permutations $\pi$: $\frac{1}{\sqrt{N!}}\sum_{\pi\in S_N} \ket{\pi}$. 
Before each query $O$, we insert a transformation 
$\ket{i}\ket{\pi}\rightarrow\ket{\pi^{-1}(i)}\ket{\pi}$
on the part of the state containing the index $i$ to be queried
and $\H_P$. After the query, we insert a transformation
$\ket{i}\ket{\pi}\rightarrow\ket{\pi(i)}\ket{\pi}$. 
At the end of algorithm, we apply the transformation
$\ket{i_1} \ldots \ket{i_K}\ket{\pi}\rightarrow \ket{\pi^{-1}(i_1)}\ldots
\ket{\pi^{-1}(i_K}\ket{\pi}$. The effect of the symmetrization 
is that, on the subspace $\ket{s}\otimes \ket{\pi}$,
the algorithm is effectively running on the input $x_1$, $\ldots$, $x_N$
with $x_{\pi(i_1)}=\ldots=x_{\pi(i_K)}=1$.

If the original algorithm $\A$ succeeds on every input $(x_1, \ldots, x_N)$
with probability at least $\epsilon$, the symmetrized algorithm
also succeeds with probability at least $\epsilon$, since its
success probability is just the average of the success probabilities of $\A$
over all $(x_1, \ldots, x_N)$ with exactly $K$ values $x_i=1$.
Next, we recast $\A$ into a different form, using a register that stores
the input $x_1, \ldots, x_N$, as in section \ref{sec:previous}. 

Let $\H_A$ be the Hilbert space on which the symmetrized version of 
$\A$ operates. Let $\H_I$ be an ${N \choose K}$-dimensional Hilbert space 
whose basis states correspond to inputs $(x_1, \ldots, x_N)$ 
with exactly $K$ values $i:x_i=1$. We transform $\A$ into a sequence of 
transformations on a Hilbert space $\H=\H_A\otimes\H_I$. 
A non-query transformation $U$ on $\H_A$ is replaced with $U\otimes I$ on $\H$.
A query is replaced by a transformation $O$ that is equal 
to $O_{x_1, \ldots, x_N}\otimes I$
on the subspace consisting of states of the form 
$\ket{s}_A\otimes\ket{x_1 \ldots x_N}_I$.
The starting state of the algorithm on Hilbert space $\H$ is 
$\ket{\varphi_0}=\ket{\psi_{start}}_A\otimes\ket{\psi_0}_I$ 
where $\ket{\psi_{start}}$ is
the starting state of $\A$ as an algorithm acting 
on $\H_A$ and $\ket{\psi_0}$ is
the uniform superposition of all basis states of $\H_I$:
\[ \ket{\psi_0}=\frac{1}{\sqrt{N\choose K}} 
\sum_{x_1, \ldots, x_N:x_1+\ldots+x_N=K}
\ket{x_1\ldots x_N} .\]
Let $\ket{\psi_t}$ be the state of the algorithm $\A$, as a sequence of transformations 
on $\H$, after the $t^{\rm th}$ query. Let $\rho_t$ be the mixed state obtained from
$\ket{\psi_t}$ by tracing out the $\H_A$ register. 

We claim that the states $\rho_t$ have a special form, 
due to our symmetrization step.

\begin{lemma}
\label{lem:sym}
The entries $(\rho_t)_{x, y}$ are the same for all $x=(x_1, \ldots, x_N)$, 
$y=(y_1, \ldots, y_N)$ with the same cardinality of the set $\{l:x_l=y_l=1\}$.
\end{lemma}

\proof
Since $\rho_t$ is independent of the way how the $\H_A\otimes \H_S$ is traced
out, we first measure $\H_S$ (in the $\ket{\pi}$ basis) and then measure $\H_A$
(arbitrarily). When measuring $\H_S$, every $\pi$ is obtained with an equal
probability. Let $\rho_{t, \pi}$ be the reduced density matrix of $\H_I$,
conditioned on the measurement of $\H_S$ giving $\pi$. Then,
\[ \rho_t =\sum_{\pi} \frac{1}{N!} \rho_{t, \pi} .\]
The entry $(\rho_{t, \pi})_{x, y}$ is the same
as the entry $(\rho_{t, id})_{\pi^{-1}(x), \pi^{-1}(y)}$ because
the symmetrization by $\pi$ maps $\pi^{-1}(x), \pi^{-1}(y)$ to $x, y$.
For every $x, y$, $x', y'$ with $|\{i:x_i=y_i=1\}|=|\{i:x'_i=y'_i=1\}|$,
there is an equal number of permutations $\pi$ mapping $\pi(x)=x'$,
$\pi(y)=y'$. Therefore, $(\rho_t)_{x, y}$ is the average of 
$(\rho_{t, id})_{x', y'}$
over all $x', y'$ with $|\{l:x_l=y_l=1\}|=|\{l:x'_l=y'_l=1\}|$.
This means that $(\rho_t)_{x, y}$ only depends on $|\{l:x_l=y_l=1\}|$. 
\qed

Any ${N \choose K}\times {N\choose K}$ matrix with 
this property shares the same eigenspaces.
Namely \cite{Knuth}, its eigenspaces are $S_0$, $S_1$, $\ldots$, $S_K$ where 
$T_0=S_0$ consists of multiples of $\ket{\psi_0}$ and, for $j>0$, 
$S_j=T_j-T_{j-1}$, with $T_j$ being the space spanned by all states 
\[ \ket{\psi_{i_1, \ldots, i_j}}=\frac{1}{\sqrt{N \choose K-j}}
\mathop{\mathop{\sum_{x_1, \ldots, x_N:}}_{x_1+\ldots+x_N=K,}}_{x_{i_1}=\ldots=x_{i_j}=1} 
\ket{x_1\ldots x_N} .\] 
Let $\tau_j$ be the completely mixed state over the subspace $S_j$.

\begin{lemma}
\label{lem:eigen}
There exist $p_{t,0}\geq 0$, $\ldots$, $p_{t, K}\geq 0$ such that 
$\rho_t=\sum_{j=0}^K p_{t, j}\tau_j$. 
\end{lemma}

\proof 
According to \cite{Knuth}, $S_0$, $\ldots$, $S_K$ are the eigenspaces of 
$\rho_t$. Therefore, $\rho_t$ is a linear combination of 
the projectors to $S_0$, $\ldots$, $S_K$.
Since $\tau_j$ is a multiple of the projector to $S_j$, we have
\[ \rho_t=\sum_{j=0}^K p_{t, j}\tau_j .\]
Since $\rho_t$ is a density matrix, it must be positive semidefinite.
This means that $p_{t,0}\geq 0$, $\ldots$, $p_{t, K}\geq 0$.
\qed

Let $q_{t, j}=p_{t, j}+p_{t, j+1}+\ldots+p_{t, K}$.
The theorem now follows from the following lemmas.

\begin{lemma}
$p_{0, 0}=1$, $p_{0, j}=0$ for $j>0$.
\end{lemma}

\proof
The state $\ket{\varphi_0}$ is just $\ket{\psi_{start}}\otimes\ket{\psi_0}$.
Tracing out $\ket{\psi_{start}}$ leaves the state 
$\rho_0=\ket{\psi_0}\bra{\psi_0}$.
\qed

\begin{lemma}
\label{lem:main1}
For all $j\in \{1, \ldots, K\}$ and all $t$,
$q_{t+1, j+1}\leq q_{t, j+1}+
\frac{4\sqrt{K}}{\sqrt{N}} q_{t, j}$
\end{lemma}

\proof
In section \ref{sec:proof1}.
\qed

\begin{lemma}
$q_{t, j}\leq {t \choose j} \left(\frac{4\sqrt{K}}{\sqrt{N}}\right)^j$.
\end{lemma}

\proof
By induction on $t$. The base case, $t=0$ follows immediately
from $p_{0, 0}=1$ and $p_{0, 1}=\ldots=p_{0, K}=0$.
For the inductive case, we have
\[ q_{t+1, j}\leq q_{t, j}+\frac{4\sqrt{K}}{\sqrt{N}} q_{t, j-1}
\leq {t \choose j} \left(\frac{4\sqrt{K}}{\sqrt{N}}\right)^j + 
\frac{4\sqrt{K}}{\sqrt{N}} {t \choose j-1} 
\left(\frac{4\sqrt{K}}{\sqrt{N}}\right)^{j-1} \]
\[ \leq \left(  {t \choose j}+{t \choose j-1} \right) 
\left (\frac{4\sqrt{K}}{\sqrt{N}}\right)^j =
{t+1\choose j} \left(\frac{4\sqrt{K}}{\sqrt{N}}\right)^j ,\]
with the first inequality following from Lemma \ref{lem:main1}
and the second inequality following from the inductive assumption.
\qed

\begin{lemma}
\label{lem:main1c}
If $t\leq 0.03 \sqrt{NK}$, then $p_{t, j}<0.65^j$ for all $j>K/2$. 
\end{lemma}

\proof
We have
\[ q_{t, j}\leq {t \choose j} \left(\frac{4\sqrt{K}}{\sqrt{N}}\right)^j <
\frac{t^j}{j!} \left(\frac{4\sqrt{K}}{\sqrt{N}}\right)^j \]
\[\leq  \frac{t^j e^j}{j^j} \left(\frac{4\sqrt{K}}{\sqrt{N}}\right)^j =
\left(\frac{4\sqrt{K} e t}{\sqrt{N} j} \right)^j ,\]
where the third inequality follows from $j!\geq (\frac{j}{e})^j$
which is a consequence of the Stirling's formula.
Let $j>K/2$ and $t\leq 0.03\sqrt{NK}$.
Then,
\[ \frac{4\sqrt{K} e t}{\sqrt{N} j} 
\leq \frac{0.12 e \sqrt{K} \sqrt{NK}}{\sqrt{N} K/2} < 0.65,\]
implying the lemma.   
\qed

\begin{lemma}
\label{lem:main2}
The success probability of $\A$ is at most 
\[ \frac{{N \choose K/2}}{{N\choose K}}+
4 \sqrt{\sum_{j=K/2+1}^K p_{T, j}} .\]
\end{lemma}

\proof
In section \ref{sec:proof2}.
\qed

To complete the proof, given the two Lemmas, 
we choose a constant $c>\sqrt[4]{0.65}=0.8979...$ and set $\epsilon=0.04$.
Then, by Lemma \ref{lem:main2}, the success probability
of $\A$ is at most 
\[ \frac{{N\choose K/2}}{{N \choose K}} 
+ 4 \sqrt{\frac{K}{2} 0.65^{K/2}} .\]
The first term is equal to
\[ \frac{{N\choose K/2}}{{N \choose K}} = 
\frac{K! (N-K)!}{(K/2)! (N-K/2)!} \leq \frac{K!}{(K/2)! (N-K)^{K/2}} 
\]
\[ = O\left(\frac{(K/e)^K}{(K/2e)^{K/2} (N-K)^{K/2}}\right) = 
O\left( \left(\frac{2K}{e(N-K)}\right)^{K/2} \right) \]
\[ = O\left(\left(\frac{2}{e}\right)^{K/2}\right) = O(0.857...^K) ,\]
with the third step following from Stirling's approximation
and the fifth step following from $K<N/2$.
The second part, $\sqrt{\frac{K}{2} 0.65^{K/2}}$ is less than $c^K/2$ if
$K$ is sufficiently large. 

It remains to prove the two lemmas.

\section{Proof of Lemma \ref{lem:main1}}
\label{sec:proof1}

%
We decompose the state $\ket{\psi_t}$ as 
$\sum_{i=0}^N a_i \ket{\psi_{t, i}}$, with 
$\ket{\psi_{t, i}}$ being the part in which the query register contains $\ket{i}$.
Because of symmetrization, we must have $|a_1|=|a_2|=\ldots=|a_N|$.
Let $\rho_{t, i}=\ket{\psi_{t, i}}\bra{\psi_{t, i}}$. 
Then, 
\begin{equation}
\label{eq-2507a}
\rho_t=\sum_{i=0}^N a^2_i \rho_{t, i}.
\end{equation}

For $i>0$, we have

\begin{Claim}
Let $i\in\{1, \ldots, N\}$.
The entry $(\rho_{t, i})_{x, y}$ only depends on $x_i, y_i$ and the cardinality
of $\{l:l\neq i, x_l=y_l=1\}$.
\end{Claim}

\proof
Similar to lemma \ref{lem:sym}.
\qed

We now describe the eigenspaces of matrices $\rho_{t, i}$. 
The proofs of some claims are postponed to section \ref{sec:eigen}.

We define the following subspaces of states.
Let $T^{i, 0}_{j}$ be the subspace spanned by all states
\[ \ket{\psi^{i,0}_{i_1, \ldots, i_j}}=\frac{1}{\sqrt{N-j-1 \choose K-j}}
\mathop{ \sum_{x:|x|=K}}_{x_{i_1}=\ldots=x_{i_j}=1, x_i=0} 
\ket{x_1\ldots x_N} \] 
and $T^{i, 1}_{j}$ be the subspace spanned by all states
\[ \ket{\psi^{i,1}_{i_1, \ldots, i_j}}=\frac{1}{\sqrt{N-j-1 \choose K-j-1}}
\mathop{ \sum_{x:|x|=K}}_{x_{i_1}=\ldots=x_{i_j}=1, x_i=1} 
\ket{x_1\ldots x_N} .\] 
Let $S^{i, 0}_{j}=T^{i, 0}_{j}\cap (T^{i, 0}_{j-1})^{\perp}$ and
$S^{i, 1}_{j}=T^{i, 1}_{j}\cap (T^{i, 1}_{j-1})^{\perp}$.
Equivalently, we can define $S^{i, 0}_j$ and $S^{i, 1}_j$ as the
subspaces spanned by the states $\ket{\tilde{\psi}^{i,0}_{i_1, \ldots, i_j}}$
and $\ket{\tilde{\psi}^{i,1}_{i_1, \ldots, i_j}}$, respectively, with
\[ \ket{\tilde{\psi}^{i,l}_{i_1, \ldots, i_j}} = P_{(T^{i, l}_{j-1})^{\perp}} 
\ket{\psi^{i,l}_{i_1, \ldots, i_j}} .\]
Let $S^i_{\alpha, \beta, j}$ be the subspace spanned by all states
\begin{equation}
\label{eq-2507}
\alpha\frac{\ket{\tilde{\psi}^{i,0}_{i_1, \ldots, i_j}}}{
\|\ket{\tilde{\psi}^{i,0}_{i_1, \ldots, i_j}}\|}+
\beta \frac{\ket{\tilde{\psi}^{i,1}_{i_1, \ldots, i_j}}}{
\|\ket{\tilde{\psi}^{i,1}_{i_1, \ldots, i_j}}\|} .
\end{equation}

\begin{Claim}
\label{claim:eigen1}
Every eigenspace of $\rho_{t,i}$ is 
a direct sum of subspaces $S^i_{\alpha, \beta, j}$ 
for some $\alpha$, $\beta$, $j$.
\end{Claim}

\proof
In section \ref{sec:eigen}.
\qed

Let $\tau^i_{\alpha, \beta, j}$ be the completely mixed state
over $S^i_{\alpha, \beta, j}$. 
Similarly to lemma \ref{lem:eigen}, we can write $\rho_{t, i}$ as
\begin{equation}
\label{eq-2507b}
 \rho_{t,i}=\sum_{(\alpha, \beta, j)\in A_{t, i}} 
p^i_{\alpha, \beta, j} \tau^{i}_{\alpha, \beta, j} ,
\end{equation}
where $(\alpha, \beta, j)$ range over some finite set $A_{t, i}$.
(This set is finite because the $\H_I$ register holding $\ket{x_1\ldots x_N}$ 
is finite dimensional and, therefore, decomposes
into a direct sum of finitely many eigenspaces.)
For every pair $(\alpha, \beta, j)\in A_{t, i}$, we 
normalize $\alpha, \beta$ by multiplying them by the same constant
so that $\alpha^2+\beta^2=1$.
Querying $x_i$ transforms this state to
\[ \rho'_{t,i}=\sum_{(\alpha, \beta, j)\in A_{t,i}} 
p^i_{\alpha, \beta, j} \tau^{i}_{\alpha, -\beta, j} ,\]
because $\ket{\tilde{\psi}^{i, l}_{i_1, \ldots, i_j}}$ is a superposition
of $\ket{x}$ with $x_i=l$ and, therefore, a query leaves
$\ket{\tilde{\psi}^{i, 0}_{i_1, \ldots, i_j}}$ unchanged and flips a phase
on $\ket{\tilde{\psi}^{i, 1}_{i_1, \ldots, i_j}}$.
If $i=0$, we have $\rho'_{t, 0}=\rho_{t, 0}$, because, if
the query register contains $\ket{0}$, the query maps any state to itself,
thus leaving $\rho_{t, 0}$ unchanged.

\begin{Claim}
\label{claim:relate}
Let $\alpha_0=\sqrt{\frac{N-K}{N-j}} 
\|\tilde{\psi}^{i, 0}_{i_1, \ldots, i_j}\|$ and
$\beta_0=\sqrt{\frac{K-j}{N-j}} \|\tilde{\psi}^{i, 1}_{i_1, \ldots, i_j}\|$.
\begin{enumerate}
\item[(i)]
$S^i_{\alpha_0, \beta_0, j}\subseteq S_j$;
\item[(ii)]
$S^i_{\beta_0, -\alpha_0, j}\subseteq S_{j+1}$.
\end{enumerate}
\end{Claim}

\proof
In section \ref{sec:eigen}.
\qed

\begin{Corollary}
\label{cor:contain}
For any $\alpha$, $\beta$, $S^i_{\alpha, \beta, j}\subseteq S_j\cup S_{j+1}$.
\end{Corollary}

\proof
We have $S_{\alpha, \beta, j}\subseteq S^{i, 0}_j \cup S^{i, 1}_j$,
since $S_{\alpha, \beta, j}$ is spanned by linear combinations of
states $\ket{\tilde{\psi}^{i, 0}_{i_1, \ldots, i_j}}$ (which belong to $S^{i, 0}_j$) and
states $\ket{\tilde{\psi}^{i, 1}_{i_1, \ldots, i_j}}$ (which belong to $S^{i, 1}_j$).
As shown in the proof of claim \ref{claim:relate} above,
\[ S^{i, 0}_j\cup S^{i, 1}_j\subseteq S_{\alpha_0, \beta_0, j}\cup 
S_{-\beta_0, \alpha_0, j} \subseteq S_j \cup S_{j+1} .\]
\qed

The next claim quantifies the overlap between $S^i_{\alpha, \beta, j}$
and $S_{j+1}$.

\begin{Claim}
\label{claim:2dim}
\[ Tr P_{S_{j+1}} \tau^{i}_{\alpha, \beta, j} = 
\frac{|\alpha\beta_0-\beta\alpha_0|^2}{\alpha_0^2 +\beta_0^2} \]
\end{Claim}

\proof
In section \ref{sec:eigen}.
\qed

To be able to use this bound, we also need to bound $\alpha_0$ and $\beta_0$.

\begin{Claim}
\label{claim:relatebound}
$\frac{\beta_0}{\sqrt{\alpha_0^2+\beta_0^2}}\leq 
\sqrt{\frac{4(K-j)}{N+3K-4j}}$.
\end{Claim}

\proof
In section \ref{sec:eigen}.
\qed

We can now complete the proof of lemma \ref{lem:main1}. 
By projecting both sides of $\rho_t=\sum_i p_{t, i}\tau_i$
to $(T_{j})^{\perp}=S_{j+1}\cup \ldots S_K$ 
and taking trace, we get 
\begin{equation}
\label{eq-project} 
Tr P_{(T_{j})^{\perp}} \rho_t = 
\sum_{j'=0}^K p_{t, j} Tr P_{(T^{j})^{\perp}} \tau_{j'}=
\sum_{j'=j}^K p_{t, j} =q_{t, j},
\end{equation}
with the second equality following because 
the states $\tau_{j'}$ are uniform mixtures over subspaces $S_{j'}$
and $S_0, \ldots, S_{j}$ are contained in $T_{j}$ while
$S_{j+1}, \ldots, S_K$ are contained in $(T_{j})^{\perp}$.
Because of equations (\ref{eq-2507a}), (\ref{eq-2507aa}) and (\ref{eq-2507b}),
this means that
\begin{equation}
\label{eq-2607b}
 q_{t, j+1} = a^2_0 Tr P_{(T_{j})^{\perp}} \rho_{t, 0} + 
\sum_{i=1}^N a^2_i \sum_{(\alpha, \beta, j') \in A_{t, i}} 
p^{i}_{\alpha, \beta, j'} 
Tr P_{(T_{j})^{\perp}}  \tau^{i}_{\alpha, \beta, j'} .
\end{equation}
Decomposing the state after the query in a similar way, we get 
\[ q_{t+1, j+1} = a^2_0 
Tr P_{(T_{j})^{\perp}}  \rho'_{t, 0} + 
 \sum_{i=1}^N a^2_i \sum_{(\alpha, \beta, j')\in A_{t, i}}   
p^{i}_{\alpha, \beta, j'} Tr P_{(T_{j})^{\perp}} 
\tau^{i}_{\alpha, -\beta, j'} .\]
By substracting the two sums and using $\rho'_{t, 0}=\rho_{t, 0}$, we get
\begin{equation}
\label{eq-2607} 
q_{t+1, j+1}-q_{t, j+1}=  \sum_{i=1}^N a^2_i 
\sum_{(\alpha, \beta, j')\in A_{t, i}} 
p^{i}_{\alpha, \beta, j'} Tr P_{(T_{j})^{\perp}} 
(\tau^{i}_{\alpha, -\beta, j'} - \tau^{i}_{\alpha, \beta, j'}) .
\end{equation}
We now claim that all the terms in this sum with $j'\neq j$ are 0.
For $j'<j$, $S_{\alpha, \beta, j'}\subseteq T_{j'+1}\subseteq T_{j}$,
implying that $Tr P_{(T_{j})^{\perp}} \tau^{i}_{\alpha, \beta, j'} = 0$
and, similarly, $Tr P_{(T_{j})^{\perp}} \tau^{i}_{\alpha, -\beta, j'} = 0$.
For $j'>j$, $S_{\alpha, \beta, j'}\subseteq S_{j'}\cup S_{j'+1} 
\subseteq (T_{j})^{\perp}$,
implying that 
\[ Tr P_{(T_{j})^{\perp}} \tau^{i}_{\alpha, \beta, j'} = 1, \mbox{~~}
Tr P_{(T_{j})^{\perp}} \tau^{i}_{\alpha, -\beta, j'} = 1 \]
and the difference of the two is 0. 
By removing those terms from (\ref{eq-2607}), we get
\begin{equation}
\label{eq-2607new} 
q_{t+1, j+1}-q_{t, j+1}=  \sum_{i=1}^N a^2_i 
\sum_{(\alpha, \beta, j)\in A_{t, i}} 
p^{i}_{\alpha, \beta, j} Tr P_{(T_{j})^{\perp}} 
(\tau^{i}_{\alpha, -\beta, j} - \tau^{i}_{\alpha, \beta, j}) .
\end{equation}

We have
\[ Tr P_{(T_{j})^{\perp}} (\tau^{i}_{\alpha, -\beta, j} - 
\tau^{i}_{\alpha, \beta, j} ) =
Tr P_{S_{j+1}} (\tau^{i}_{\alpha, -\beta, j} - 
\tau^{i}_{\alpha, \beta, j} ) \]
\[ =
\frac{|\alpha\beta_0+\beta\alpha_0|^2}{\alpha_0^2 +\beta_0^2} - 
\frac{|\alpha\beta_0-\beta\alpha_0|^2}{\alpha_0^2 +\beta_0^2}  ,\]
with the first equality following from Corollary \ref{cor:contain},
$S_{j}\subseteq T_{j}$ and $S_{j+1}\subseteq (T_{j})^{\perp}$
and the second equality following from Claim \ref{claim:2dim}.
This is at most
\[ 4 \frac{|\alpha \beta \alpha_0 \beta_0|}{\alpha_0^2 +\beta_0^2} 
\leq 2 \frac{\alpha_0 \beta_0}{\alpha_0^2 +\beta_0^2} \] 
\[ =2 \frac{\alpha_0}{\sqrt{\alpha_0^2+\beta_0^2}}  
\frac{\beta_0}{\sqrt{\alpha_0^2+\beta_0^2}}
\leq 2 \sqrt{\frac{4(K-j)}{N+3K-4j}} \leq 2 \sqrt{\frac{4K}{N}} ,\]
with the first inequality following from 
$|\alpha\beta|\leq \frac{|\alpha|^2+|\beta|^2}{2}=\frac{1}{2}$
and the second inequality following from Claim \ref{claim:relatebound}
and $\frac{\alpha_0}{\sqrt{\alpha_0^2+\beta_0^2}}\leq 1$.
Together with equation (\ref{eq-2607}), this means
\begin{equation}
\label{eq-almost} 
q_{t+1, j+1}-q_{t, j+1} \leq \frac{4\sqrt{K}}{\sqrt{N}} 
\sum_{i=1}^N a_i^2 \sum_{(\alpha, \beta, j)\in A_{t, i}} 
p^{i}_{\alpha, \beta, j} 
\end{equation}
Similarly to equation (\ref{eq-project}) we have
\[ p_{t, j+1}+p_{t, j} = Tr P_{(S_j \cup S_{j+1})} \rho_t .\]
We can then express the right hand side similarly to
equation (\ref{eq-2607b}), as a sum of terms
$p^0_{j'} Tr P_{(S_j \cup S_{j+1})} \tau_{j'}$ and 
$p^i_{\alpha, \beta, j'} Tr P_{(S_j \cup S_{j+1})} \tau^i_{\alpha, \beta, j'}$.
Since $S^i_{\alpha, \beta, j}\subseteq S_j \cup S_{j+1}$
(by corollary \ref{cor:contain}), we have
$Tr P_{(S_{j}\cup S_{j+1})} \tau^{i}_{\alpha, \beta, j}=1$.
This means that 
\[ p_{t, j+1}+p_{t, j} \geq \sum_{i=1}^N a_i^2 
\sum_{(\alpha, \beta, j)\in A_{t, i}} 
p^{i}_{\alpha, \beta, j} .\]
Together with equation (\ref{eq-almost}), this implies
\[ q_{t+1, j+1}-q_{t, j+1} \leq
\frac{4\sqrt{K}}{\sqrt{N}} (p_{t, j}+p_{t, j+1}) \leq 
\frac{4\sqrt{K}}{\sqrt{N}} \sum_{j'=j}^K p_{t, j'} =
\frac{4\sqrt{K}}{\sqrt{N}} q_{t, j} .\]
\qed

\section{Proof of Lemma \ref{lem:main2}}
\label{sec:proof2}

We start with the case, when $p_{T, K/2+1}=\ldots=p_{T, K}=0$.

\begin{lemma}
\label{lem:modifiedprob}
If $p_{T, K/2+1}=\ldots=p_{T, K}=0$, the success probability of $\A$ is
at most $\frac{{N\choose K/2}}{{N \choose K}}$.
\end{lemma}

\proof
Let $\ket{\psi}$ be the final state. 
The state of $\H_I$ register lies in $T_{K/2}$, which is
a ${N \choose K/2}$ dimensional space. 
Therefore, there is a Schmidt decomposition for $\ket{\psi}$
with at most ${N \choose K/2}$ terms.
This means that the state of $\H_A$ lies in a ${N \choose K/2}$
subspace of $\H_A\otimes H_S$.

We express the final state as 
\[ \ket{\psi}= \sum_{x:|x|=K} \frac{1}{\sqrt{{N \choose K}}} 
\ket{\psi_x} \ket{x} .\]
We can think of $\ket{\psi_x}$ as a quantum encoding for $x$
and the final measurement as a decoding procedure that
takes $\ket{\psi_x}$ and produces a guess for $x$.
The probability that algorithm $\A$ succeeds is then
equal to the average success probability of the encoding. 
We now use
\begin{theorem}
\label{thm:nayak}
\cite{Nayak}
For any encoding $\ket{\psi_x}$ of $M$ classical values in
by quantum states in $d$ dimensions, the probability of
success is at most $\frac{d}{M}$.
\end{theorem}

In our case, $M={N \choose K}$ and $d={N\choose K/2}$ because the
states $\ket{\psi}$ all lie in a ${N \choose K/2}$-dimensional
subspace of $\H_A\otimes \H_S$. Therefore, Theorem \ref{thm:nayak}
implies Lemma \ref{lem:modifiedprob}.
\qed

We decompose the state $\ket{\psi_T}$ as 
$\sqrt{1-\delta}\ket{\psi'_T}+\sqrt{\delta}\ket{\psi''_T}$
where $\ket{\psi'_T}$ is in the subspace 
$\H_A \otimes \cup_{j=0}^{K/2} S_j$ and 
$\ket{\psi''_T}$ is in $\H_A \otimes \cup_{j=K/2+1}^{K} S_j$.
We have 
\[ \delta=\sum_{j=K/2+1}^K p_{T, j} .\]
The success probability of $\A$ is the probability that, if we measure both
the register of $\H_A$ containing the result of the computation and $\H_I$,
then, we get $i_1, \ldots, i_K$ and $x_1, \ldots, x_N$ such that 
$x_{i_1}=\ldots=x_{i_K}=1$. 

Consider the probability of getting $i_1, \ldots, i_K$ 
and $x_1, \ldots, x_N$ such that 
$x_{i_1}=\ldots=x_{i_K}=1$, when measuring $\ket{\psi'_T}$ (instead of
$\ket{\psi_T}$). By Lemma \ref{lem:modifiedprob}, this probability is
at most $\frac{{N\choose K/2}}{{N \choose K}}$.
We have 
\[ \|\psi_T -\psi'_T\| \leq (1-\sqrt{1-\delta^2}) 
\|\psi'_T\| + \sqrt{\delta} \|\psi''_T\| =
(1-\sqrt{1-\delta^2}) + \sqrt{\delta} \leq 2 \sqrt{\delta} .\]
We now apply

\begin{lemma}
\label{lem:bv}
\cite{BV}
For any states $\ket{\psi}$ and $\ket{\psi'}$ and any measurement $M$,
the variational distance between the 
probability distributions obtained by applying $M$ to $\ket{\psi}$
and $\ket{\psi'}$ is at most $2\|\psi-\psi'\|$.
\end{lemma}

By Lemma \ref{lem:bv}, the probabilities of getting $i_1, \ldots, i_K$ 
and $x_1, \ldots, x_N$ such that $x_{i_1}=\ldots=x_{i_K}=1$, 
when measuring $\ket{\psi_T}$ and $\ket{\psi'_T}$
differ by at most $4\sqrt{\delta}=4\sqrt{\sum_{j=K/2+1}^K p_{T, j}}$.
Therefore, the success probability of $\A$ is at most
\[ \frac{{N\choose K/2}}{{N \choose K}} + 4\sqrt{\sum_{j=K/2+1}^K p_{T, j}} .\]

\section{Structure of the eigenspaces of $\rho_{t, i}$}
\label{sec:eigen}

In this section, we prove claims \ref{claim:eigen1}, \ref{claim:relate},
\ref{claim:2dim} and \ref{claim:relatebound} describing
the structure of the eigenspaces of $\rho_{t, i}$.

\proof [of Claim \ref{claim:eigen1}]
%
We rearrange the rows and the columns of $\rho_{t, i}$ so that all rows
and columns corresponding to $\ket{x_1\ldots x_N}$ with $x_i=0$ are before
the rows and the columns corresponding to $\ket{x_1\ldots x_N}$ with $x_i=1$. 
We then express $\rho_{t, i}$ as
\[ \rho_{t, i}= \left(
\begin{array}{cc} A & B \\
C & D \end{array} \right) ,\]
with $A$ being a ${N-1 \choose K}\times {N-1\choose K}$ square matrix
indexed by $\ket{x_1\ldots x_N}$ with $x_i=0$, 
$D$ being a ${N-1 \choose K-1}\times {N-1\choose K-1}$ square matrix
indexed by $\ket{x_1\ldots x_N}$ with $x_i=1$ and $B$ and $C$ 
being rectangular matrices with rows (columns) indexed by 
$\ket{x_1\ldots x_N}$ with $x_i=0$ and columns (rows) indexed by
$\ket{x_1\ldots x_N}$ with $x_i=1$.

We claim that 
\[ \rho_{t, i} \ket{\tilde{\psi}^{i,0}_{i_1, \ldots, i_j}} 
= a_{11} \ket{\tilde{\psi}^{i,0}_{i_1, \ldots, i_j}}+
a_{12} \ket{\tilde{\psi}^{i,1}_{i_1, \ldots, i_j}} ,\]
\begin{equation} 
\label{eq-rem0}
\rho_{t, i} \ket{\tilde{\psi}^{i,1}_{i_1, \ldots, i_j}} 
= a_{21} \ket{\tilde{\psi}^{i,0}_{i_1, \ldots, i_j}}+
a_{22} \ket{\tilde{\psi}^{i,1}_{i_1, \ldots, i_j}} ,
\end{equation}
where $a_{11}$, $a_{12}$, $a_{21}$, $a_{22}$ are independent of $i_1, \ldots, i_j$.
To prove that, we first note that $A$ and $D$ are matrices where $A_{xy}$ and $D_{xy}$ only
depends on $|\{ t:x_t=y_t\}|$. Therefore, the results of Knuth\cite{Knuth} about 
eigenspaces of such matrices apply. This means that $S^{i, 0}_j$ an $S^{i, 1}_j$
are eigenspaces for $A$ and $D$, respectively, and
\[ A \ket{\tilde{\psi}^{i,0}_{i_1, \ldots, i_j}} = a_{11} \ket{\tilde{\psi}^{i,0}_{i_1, \ldots, i_j}},\]
\[ D \ket{\tilde{\psi}^{i,1}_{i_1, \ldots, i_j}} = a_{22} \ket{\tilde{\psi}^{i,1}_{i_1, \ldots, i_j}},\]
where $a_{11}$ and $a_{22}$ are the eigenvalues of $A$ and $D$ for 
the eigenspaces $S^{i, 0}_j$ and $S^{i, 1}_j$.
It remains to prove that
\begin{equation}
\label{eq-rem1} 
B \ket{\tilde{\psi}^{i,0}_{i_1, \ldots, i_j}} = a_{12} \ket{\tilde{\psi}^{i,1}_{i_1, \ldots, i_j}},
\end{equation}
\begin{equation}
\label{eq-rem2}
C \ket{\tilde{\psi}^{i,1}_{i_1, \ldots, i_j}} = a_{21} \ket{\tilde{\psi}^{i,0}_{i_1, \ldots, i_j}}.
\end{equation}
Let $M$ be a rectangular matrix, with entries indexed by $x, y$,
with $|x|=|y|=K$ and $x_i=1$ and $y_i=0$.
The entries of $M$ are $M_{xy}=1$ if $x$ and $y$ differ in two places, with $x_i=1$, $y_i=0$
and $x_l=0$, $y_l=1$ for some $l\neq i$ and $M_{xy}=0$ otherwise.
We claim 
\begin{equation}
\label{eq-rem3}
M\ket{\tilde{\psi}^{i,0}_{i_1, \ldots, i_j}}=c \ket{\tilde{\psi}^{i,1}_{i_1, \ldots, i_j}}
\end{equation}
for some $c$ that may depend on $N, k$ and $j$ but not on $i_1, \ldots, i_j$.
To prove that, we need to prove two things.
First, 
\begin{equation}
\label{eq-rem4}
M\ket{\psi^{i,0}_{i_1, \ldots, i_j}}=c \ket{\psi^{i,1}_{i_1, \ldots, i_j}}.
\end{equation}
This follows by
\[ M\ket{\psi^{i,0}_{i_1, \ldots, i_j}} = \frac{1}{\sqrt{N-j-1 \choose k-j}} 
\mathop{\sum_{x:x_{i_1}=\ldots=x_{i_j}=1,}}_{x_i=0} M \ket{x}  \] 
\[=\frac{1}{\sqrt{N-j-1 \choose K-j}} 
\mathop{\sum_{x:x_{i_1}=\ldots=x_{i_j}=1}}_{x_i=0} \sum_{l:x_l=1} 
\ket{x_1 \ldots x_{l-1} 0 x_{l+1} \ldots x_{i-1} 1 x_{i+1} \ldots x_N} \]
\[ =\frac{1}{\sqrt{N-j-1 \choose K-j}} (N-K) 
\mathop{\sum_{y:y_{i_1}=\ldots=y_{i_j}=1}}_{y_i=1} \ket{y}=
\sqrt{(K-j)(N-K)} \ket{\psi^{i,1}_{i_1, \ldots, i_j}} .\] 

Second, $M(T^{i,0}_j)\subseteq T^{i,1}_j$ and 
$M (T^{i, 0}_j)^{\perp} \subseteq (T^{i,1}_j)^{\perp}$.
The first statement is immediately follows from equation (\ref{eq-rem4}),
because the subspaces $T^{i,0}_j$, $T^{i, 1}_j$ are spanned by the states 
$\ket{\psi^{i,0}_{i_1, \ldots, i_j}}$ and $\ket{\psi^{i,1}_{i_1, \ldots, i_j}}$,
respectively. 
To prove the second statement, let $\ket{\psi}\in (T^{i, 0}_j)^{\perp}$, 
$\ket{\psi}=\sum_{x} a_{x}\ket{x}$. 
We would like to prove $M\ket{\psi}\in (T^{i,1}_j)^{\perp}$.
This is equivalent to $\bra{\psi^{i,1}_{i_1, \ldots, i_j}}M\ket{\psi}=0$
for all $i_1, \ldots, i_j$. 
We have 
\[ \bra{\psi^{i,1}_{i_1, \ldots, i_j}}M\ket{\psi}= \frac{1}{\sqrt{N-j-1 \choose K-j-1}} 
\sum_{y:y_{i_1}=\ldots=y_{i_j}=1} \bra{y}M\ket{\psi}\]
\[ =\frac{1}{\sqrt{N-j-1 \choose K-j-1}} 
\mathop{\sum_{x:x_{i_1}=\ldots=x_{i_j}=1,}}_{x_i=0} 
\mathop{\sum_{l:x_l=1,}}_{l\notin \{i_1, \ldots, i_j\}} a_x  \] 
\[= \frac{1}{\sqrt{N-j-1 \choose K-j-1}} 
(K-j) \sum_{x:x_{i_1}=\ldots=x_{i_j}=1} a_x = 0.\]
The first equality follows by writing out $\bra{\psi^{i,1}_{i_1, \ldots, i_j}}$,
the second equality follows by writing out $M$. The third equality follows because,
for every $x$ with $|x|=K$ and $x_{i_1}=\ldots=x_{i_j}=1$, 
there are $K-j$ more $l\in[N]$ satisfying $x_l=1$.
The fourth equality follows because $\sum_{x:x_{i_1}=\ldots=x_{i_j}=1} a_x$ is a constant
times $\bra{\psi^{i,0}_{i_1, \ldots, i_j}}\psi\rket$ and 
$\bra{\psi^{i,0}_{i_1, \ldots, i_j}}\psi\rket=0$, because $\ket{\psi}\in (T^{i, 0}_j)^{\perp}$.

Furthermore, $BM$ is an ${N-1 \choose K}\times {N-1\choose K}$ matrix, with $(BM)_{x, y}$ only
depending on $|\{l:x_l = y_l =1\}|$. Therefore, $S^{i,1}_j$ is an eigenspace of $BM$
and, since $\ket{\tilde{\psi}^{i,1}_{i_1, \ldots, i_j}}\in S^{i,1}_j$, we have  
\[ BM \ket{\tilde{\psi}^{i,1}_{i_1, \ldots, i_j}} =
\lambda \ket{\tilde{\psi}^{i,1}_{i_1, \ldots, i_j}} \]
for an eigenvalue $\lambda$ independent of $i_1, \ldots, i_j$.
Together with equation (\ref{eq-rem3}), this implies equation (\ref{eq-rem1}) with
$a_{12}=\lambda/j$.

Equation (\ref{eq-rem2}) follows by proving   
\[ M^T \ket{\tilde{\psi}^{i,1}_{i_1, \ldots, i_j}}=c \ket{\tilde{\psi}^{i,0}_{i_1, \ldots, i_j}} \]
and 
\[ CM^T \ket{\tilde{\psi}^{i,0}_{i_1, \ldots, i_j}} =
\lambda \ket{\tilde{\psi}^{i,0}_{i_1, \ldots, i_j}} ,\]
in a similar way.

We now diagonalize the matrix 
\[ M'=\left(\begin{array}{cc} a_{11} & a_{12} \\a_{21} & a_{22} \end{array} \right) .\]
It has two eigenvectors: $\left(\begin{array}{c} \alpha_1 \\ \beta_1 \end{array}\right)$ 
and $\left(\begin{array}{c} \alpha_2 \\ \beta_2 \end{array}\right)$. 
Equation (\ref{eq-rem0}) implies that, for any $i_1, \ldots, i_j$,
\[ \alpha_1 \ket{\tilde{\psi}^{i,0}_{i_1, \ldots, i_j}} + 
\beta_1 \ket{\tilde{\psi}^{i,1}_{i_1, \ldots, i_j}} \]
is an eigenvector of $M$ with the same eigenvalue $\lambda$.
Therefore, $S_{\alpha_1, \beta_1, i}$ is an eigenspace of $M$.
Similarly, $S_{\alpha_2, \beta_2, i}$ is an eigenspace of $M$.
Vectors $\alpha_1\ket{\tilde{\psi}^{i,0}_{i_1, \ldots, i_j}} + 
\beta_1 \ket{\tilde{\psi}^{i,1}_{i_1, \ldots, i_j}}$ and
$\alpha_2\ket{\tilde{\psi}^{i,0}_{i_1, \ldots, i_j}} + 
\beta_2 \ket{\tilde{\psi}^{i,1}_{i_1, \ldots, i_j}}$ together
span the same space as vectors $\ket{\tilde{\psi}^{i,0}_{i_1, \ldots, i_j}}$
and $\ket{\tilde{\psi}^{i,1}_{i_1, \ldots, i_j}}$.
Since vectors $\ket{\tilde{\psi}^{i,l}_{i_1, \ldots, i_j}}$
span $S^{i, l}_j$, this means that 
\[ S^{i,0}_j\cup S^{i,1}_j \subseteq S_{\alpha_1, \beta_1, i} \cup
S_{\alpha_2, \beta_2, i}. \]
Therefore, repeating this argument for every $i$ gives
a collection of eigenspaces that span the entire state space for $\H_I$.
This means that any eigenspace of $M$ is a direct sum of some of eigenspaces
$S_{\alpha, \beta, i}$.
\qed

\proof [of Claim \ref{claim:relate}]
For part (i), consider the states $\ket{\psi_{i_1, \ldots, i_j}}$ 
spanning $T_j$. 
We have
\begin{equation}
\label{eq-new} 
\ket{\psi_{i_1, \ldots, i_j}}=\sqrt{\frac{N-k}{N-j}} \ket{\psi^{i,0}_{i_1, \ldots, i_j}}+
\sqrt{\frac{K-j}{N-j}} \ket{\psi^{i,1}_{i_1, \ldots, i_j}} 
\end{equation}
because a $\frac{N-K}{N-j}$ fraction of the states $\ket{x_1\ldots x_N}$ with
$|x|=K$ and $x_{i_1}=\ldots=x_{i_j}=1$ have $x_i=0$ and the rest have $x_i=1$.
The projection of these states to $(T^{i,0}_{j-1}\cup T^{i, 1}_{j-1})^{\perp}$
are
\[
\sqrt{\frac{N-K}{N-j}} \ket{\tilde{\psi}^{i,0}_{i_1, \ldots, i_j}}+
\sqrt{\frac{K-j}{N-j}} \ket{\tilde{\psi}^{i,1}_{i_1, \ldots, i_j}} 
\]
which, by equation (\ref{eq-2507}) are exactly the states spanning
$S^i_{\alpha_0, \beta_0, j}$.
Furthermore, we claim that 
\begin{equation}
\label{eq-2707} T_{j-1}\subseteq T^{i, 0}_{j-1}\cup T^{i, 1}_{j-1} \subseteq T_j. 
\end{equation}
The first containment is true because $T_{j-1}$ is spanned by 
the states $\ket{\psi_{i_1, \ldots, i_{j-1}}}$
which either belong to $T^{i,1}_{j-2}\subseteq T^{i,1}_{j-1}$
(if one of $i_1, \ldots, i_{j-1}$ is equal to $i$)
or are a linear combination of states $\ket{\psi^{i,0}_{i_1, \ldots, i_{j-1}}}$
and $\ket{\psi^{i,1}_{i_1, \ldots, i_{j-1}}}$ which belong to
$T^{i,0}_{j-1}$ and $T^{i,1}_{j-1}$.
The second containment follows because the
states $\ket{\psi^{i, 1}_{i_1, \ldots, i_{j-1}}}$ spanning
$T^{i,1}_{j-1}$ are the same as the states 
$\ket{\psi_{i, i_1, \ldots, i_{j-1}}}$ which belong to $T_j$ and
the states $\ket{\psi^{i, 0}_{i_1, \ldots, i_{j-1}}}$ spanning
$T^{i,0}_{j-1}$ can be expressed as linear combinations
of $\ket{\psi_{i_1, \ldots, i_{j-1}}}$ and
$\ket{\psi_{i, i_1, \ldots, i_{j-1}}}$ which both belong to $T_j$.

The first part of (\ref{eq-2707}) now implies 
\[ S^i_{\alpha_0, \beta_0, j}\subseteq (T^{i, 0}_{j-1} \cup T^{i, 1}_{j-1})^{\perp} \subseteq
(T_{j-1})^{\perp} .\]
We also have
$S^i_{\alpha_0, \beta_0, j}\subseteq T_j$,
because, $S^i_{\alpha_0, \beta_0, j}$
is spanned by the states
\[ P_{(T^{i, 0}_{j-1} \cup T^{i, 1}_{j-1})^{\perp}} \ket{\psi_{i_1, \ldots, i_j}} =
\ket{\psi_{i_1, \ldots, i_j}} - P_{T^{i, 0}_{j-1} \cup T^{i, 1}_{j-1}} 
\ket{\psi_{i_1, \ldots, i_j}} \]
and $\ket{\psi_{i_1, \ldots, i_j}}$ belongs to $T_j$ by the definition of $T_j$ and 
$P_{T^{i, 0}_{j-1} \cup T^{i, 1}_{j-1}} \ket{\psi_{i_1, \ldots, i_j}}$ belongs to
$T_j$ because of the second part of (\ref{eq-2707}).
Therefore, 
$S^i_{\alpha_0, \beta_0, j} \subseteq T_j \cap (T_{j-1})^{\perp}=S_j$.

For the part (ii), 
we have
\[ S^i_{\alpha_0, \beta_0, j}\subseteq S^{i,0}_{j}\cup S^{i, 1}_j
\subseteq T^{i, 0}_j \cup T^{i, 1}_j \subseteq T_{j+1} ,\]
where the first containment is true because $S^i_{\alpha_0, \beta_0, j}$
is spanned by linear combinations of vectors
$\ket{\tilde{\psi}^{i,0}_{i_1, \ldots, i_j}}$ (which belong to $S^{i, 0}_j$)
and vectors $\ket{\tilde{\psi}^{i,1}_{i_1, \ldots, i_j}}$ 
(which belong to $S^{i, 1}_j$) and the last containment is
true because of the second part of equation (\ref{eq-2707}).

Let
\begin{equation}
\label{eq-2707a} 
\ket{\psi}=\beta_0 \frac{\ket{\tilde{\psi}^{i,0}_{i_1, \ldots, i_j}}}{
\|\ket{\tilde{\psi}^{i,0}_{i_1, \ldots, i_j}}\| }
-\alpha_0 \frac{\ket{\tilde{\psi}^{i,1}_{i_1, \ldots, i_j}}}{
\|\ket{\tilde{\psi}^{i,1}_{i_1, \ldots, i_j}}\|} 
\end{equation}
be one of the vectors spanning $S^i_{\beta_0, -\alpha_0, j}$.
To prove that $\ket{\psi}$ is in $S_{j+1}=T_{j+1} - T_j$, it remains to
prove that $\ket{\psi}$ is orthogonal to $T_{j}$.
This is equivalent to proving that $\ket{\psi}$ is orthogonal to
every of the vectors $\ket{\psi_{i'_1, \ldots, i'_j}}$ spanning $T_j$.

\noindent
{\bf Case 1.}
$\{i'_1, \ldots, i'_j\} = \{i_1, \ldots, i_j \}$.

Since $\ket{\psi}$ belongs to $(T^{i,0}_{j-1}\cup T^{i, 1}_{j-1})^{\perp}$,
it suffices to prove that $\ket{\psi}$ is orthogonal to the projection
of  $\ket{\psi_{i_1, \ldots, i_j}}$ to $(T^{i,0}_{j-1}\cup T^{i, 1}_{j-1})^{\perp}$
which, by discussion after the equation (\ref{eq-new}), 
is equal to 
\begin{equation}
\label{eq-2707b} 
\alpha_0 \frac{\ket{\tilde{\psi}^{i,0}_{i_1, \ldots, i_j}}}{
\|\ket{\tilde{\psi}^{i,0}_{i_1, \ldots, i_j}}\| }
+\beta_0 \frac{\ket{\tilde{\psi}^{i,1}_{i_1, \ldots, i_j}}}{
\|\ket{\tilde{\psi}^{i,1}_{i_1, \ldots, i_j}}\|} .
\end{equation}
From equations (\ref{eq-2707a}) and (\ref{eq-2707b}), we see that the
inner product of the two states is $\alpha_0 \beta_0 -\beta_0 \alpha_0=0$.

\noindent
{\bf Case 2.}
$\{i'_1, \ldots, i'_j\} \neq \{i_1, \ldots, i_j \}$ 
but one of $i'_1, \ldots, i'_j$ is equal to $i$.

For simplicity, assume $i=i'_j$. Then,
$\ket{\psi_{i'_1, \ldots, i'_j}}$ is the same 
as $\ket{\psi^{i, 1}_{i'_1, \ldots, i'_{j-1}}}$
which belongs to $T^{i,1}_{j-1}$. By definition of $S^i_{\alpha, \beta, j}$,
the vector $\ket{\psi}$ belongs to $(T^{i,0}_{j-1}\cup T^{i,1}_{j-1})^{\perp}$
and is therefore orthogonal to $\ket{\psi^{i, 1}_{i'_1, \ldots, i'_{j-1}}}$. 

\noindent
{\bf Case 3.}
$\{i'_1, \ldots, i'_j\} \neq \{i_1, \ldots, i_j \}$ 
and none of $i'_1, \ldots, i'_j$ is equal to $i$.

One of $i'_1, \ldots, i'_j$ must be not in $\{i_1, \ldots, i_j \}$.
For simplicity, assume it is $i'_j$. We have
\[ \ket{\psi_{i'_1, \ldots, i'_{j-1}}}= 
\sum_{i'\notin\{i'_1, \ldots, i'_{j-1}\}} 
\ket{\psi_{i'_1, \ldots, i'_{j-1}, i'}} .\]
Also, $\lbra \psi_{i'_1, \ldots, i'_{j-1}} \ket{\psi}=0$,
because $\ket{\psi_{i'_1, \ldots, i'_{j-1}}}$ is in 
$T^{i,0}_{j-1}\cup T^{i,1}_{j-1}$.
As proven in the previous case, 
$\lbra \psi_{i'_1, \ldots, i'_{j-1},i} \ket{\psi}=0$.
We therefore have
\begin{equation}
\label{eq-new1} 
\sum_{i'\notin\{i'_1, \ldots, i'_{j-1}, i\}} 
\bra{\psi_{i'_1, \ldots, i'_{j-1}, i'}} \psi\rket
=0 .
\end{equation}
By symmetry, the inner product $\bra{\psi_{i'_1, \ldots, i'_{j-1}, i'}} \psi\rket$
is the same for every $i'\notin\{i'_1, \ldots, i'_{j-1}, i\}$.
Therefore, equation (\ref{eq-new1}) means  
\[ \bra{\psi_{i'_1, \ldots, i'_{j-1}, i'}} \psi\rket
=0 \]
for every $i'\notin\{i'_1, \ldots, i'_{j-1}, i\}$.
\qed

\proof [of Claim \ref{claim:2dim}]
$\tau^i_{\alpha, \beta, j}$ is a mixture of states $\ket{\psi}$
from the subspace $S^i_{\alpha, \beta, j}$.
We prove the claim by showing that, 
for any of those states $\ket{\psi}$, the squared norm
of its projection to $S_{j+1}$ is equal to the
right hand side of claim \ref{claim:2dim}.
Since $\ket{\psi}\in S^i_{\alpha, \beta, j}$ we can write it
as 
\[ \ket{\psi} = \sum_{i_1, \ldots, i_{j}} a_{i_1, \ldots, i_{j}}
(\alpha \ket{\tilde{\psi}^{i, 0}_{i_1, \ldots, i_{j}}} +
\beta \ket{\tilde{\psi}^{i, 1}_{i_1, \ldots, i_{j}}} ) \]
for some $a_{i_1, \ldots, i_{j}}$. Let
\[ \ket{\psi^+}= \sum_{i_1, \ldots, i_{j}} a_{i_1, \ldots, i_{j}}
(\beta_0 \ket{\tilde{\psi}^{i, 0}_{i_1, \ldots, i_{j}}} -
\alpha_0 \ket{\tilde{\psi}^{i, 1}_{i_1, \ldots, i_{j}}} ) ,\]
\[ \ket{\psi^-}= \sum_{i_1, \ldots, i_{j}} a_{i_1, \ldots, i_{j}}
(\alpha_0 \ket{\tilde{\psi}^{i, 0}_{i_1, \ldots, i_{j}}} +
\beta_0 \ket{\tilde{\psi}^{i, 1}_{i_1, \ldots, i_{j}}} ) .\]
Then, $\ket{\psi}$ is a linear combination of $\ket{\psi^+}$
which belongs to $S^i_{\beta_0, -\alpha_0, j}\subset S_{j+1}$
(by Claim \ref{claim:relate}) and $\ket{\psi^-}$
which belongs to $S^i_{\alpha_0, \beta_0, j}\subseteq S_j$.
Moreover, all three states are linear combinations of
$\ket{\psi^0}$, $\ket{\psi^1}$ defined by
\[ \ket{\psi^l}=\sum_{i_1, \ldots, i_{j}} a_{i_1, \ldots, i_{j}} 
\ket{\tilde{\psi}^{i, l}_{i_1, \ldots, i_{j}}}.\]
We have 
\[ \ket{\psi}=\alpha\ket{\psi^0}+\beta\ket{\psi^1}, \]
\[ \ket{\psi^+}=\beta_0\ket{\psi^0}-\alpha_0\ket{\psi^1}, \]
\[ \ket{\psi^-}=\alpha_0\ket{\psi^0}+\beta_0\ket{\psi^1} .\]
Since $\ket{\psi^+}$ and $\ket{\psi^-}$ belong to subspaces 
$S_{j+1}$ and $S_j$ which are orthogonal, we must have
$\lbra \psi^+ \ket{\psi^-}=0$. This means
\[ \alpha_0\beta_0 \|\psi^0\|^2 -\beta_0 \alpha_0 \|\psi^1\|^2 = 0 .\]
By dividing the equation by $\alpha_0\beta_0$, we get 
$\|\psi^0\|^2=\|\psi^1\|^2$ and $\|\psi^0\|=\|\psi^1\|$.
Since $\|\psi\|=1$, this means that $\|\psi^0\|=\|\psi^1\|
=\frac{1}{\sqrt{\alpha^2+\beta^2}}=1$.

Since $\ket{\psi}$ lies in the subspace spanned by 
$\ket{\psi^+}$ which belongs to $S_{j+1}$
and $\ket{\psi^-}$ which belongs to $S_j$,
the norm of the projection of $\ket{\psi}$ to $S_{j+1}$
is equal to $\frac{|\lbra\psi|\psi^+\rket|}{\|\psi^+\|}$.
By expressing $\ket{\psi}$, $\ket{\psi^+}$ in terms of
$\ket{\psi^0}$, $\ket{\psi^1}$, we get
\[  \frac{|\lbra\psi|\psi^+\rket|}{\|\psi^+\|} =
\frac{\alpha\beta_0 \|\psi^0\|^2 -\alpha_0 \beta \|\psi^1\|^2}{
\sqrt{\beta_0^2 \|\psi^0\|^2+ \alpha_0^2 \|\psi^0\|^2 }} =
\frac{|\alpha\beta_0 -\alpha_0 \beta|}{\sqrt{\alpha_0^2+\beta_0^2}} ,\]
proving the claim.
\qed

\proof [of Claim \ref{claim:relatebound}]
We will prove
$\|\tilde{\psi}^{i, 0}_{i_1, \ldots, i_j}\| 
\geq \frac{1}{2} \|\tilde{\psi}^{i, 1}_{i_1, \ldots, i_j}\|$,
because that means 
\[ \alpha_0 = \frac{\sqrt{N-K}}{\sqrt{N-j}} 
\|\tilde{\psi}^{i, 0}_{i_1, \ldots, i_j}\|
\geq  \frac{1}{2} 
\frac{\sqrt{N-K}}{\sqrt{K-j}} \frac{\sqrt{K-j}}{\sqrt{N-j}}
\|\tilde{\psi}^{i, 1}_{i_1, \ldots, i_j}\|
= \frac{\sqrt{N-K}}{2\sqrt{K-j}} \beta_0 \]
and
\[ \frac{\beta_0}{\sqrt{\alpha_0^2+\beta_0^2}} \leq
\frac{\beta_0}{\sqrt{\frac{N-K}{4(K-j)}\beta_0^2 +\beta_0^2}} =
\frac{1}{\sqrt{1+\frac{N-K}{4(K-j)}}} = 
\frac{\sqrt{4(K-j)}}{\sqrt{N+3K-4j}} .\]

To prove $\|\tilde{\psi}^{i, 0}_{i_1, \ldots, i_j}\| 
\geq \|\tilde{\psi}^{i, 1}_{i_1, \ldots, i_j}\|$,
we calculate the vector
\[ \ket{\tilde{\psi}^{i, 0}_{i_1, \ldots, i_j}}=P_{(T^{i, 0}_{j-1})^{\perp}} 
\ket{\psi^{i, 0}_{i_1, \ldots, i_j}} .\]
Both vector $\ket{\psi^{i, 0}_{i_1, \ldots, i_j}}$ and 
subspace $T^{i, 0}_{j-1}$ are fixed by 
\[ U_\pi \ket{x}=\ket{x_{\pi(1)}\ldots x_{\pi(N)}} \]
for any permutation $\pi$ that fixes $i$ and
maps $\{i_1, \ldots, i_j\}$ to itself.
This means that $\ket{\tilde{\psi}^{i, 0}_{i_1, \ldots, i_j}}$
is fixed by any such $U_{\pi}$ as well.
Therefore, the amplitude of $\ket{x}$, $|x|=K$, $x_i=0$ in 
$\ket{\tilde{\psi}^{i, 0}_{i_1, \ldots, i_j}}$ only depends on
$|\{i_1, \ldots, i_j\}\cap \{t:x_t=1\}|$.
This means $\ket{\tilde{\psi}^{i, 0}_{i_1, \ldots, i_j}}$ is 
of the form
\[ \ket{\psi_0}=\sum_{m=0}^j \alpha_m \mathop{\sum_{x:|x|=K, x_i=0}}_{
|\{i_1, \ldots, i_j\}\cap \{t:x_t=1\}| = m} \ket{x} .\]
To simplify the following calculations, we multiply 
$\alpha_0$, $\ldots$, $\alpha_j$ by the same constant 
so that $\alpha_j=1/\sqrt{{N-j-1 \choose K-j}}$. 
Then, 
$\ket{\tilde{\psi}^{i, 0}_{i_1, \ldots, i_j}}$
remains a multiple of $\ket{\psi_0}$ but may no longer be
equal to $\ket{\psi_0}$. 

$\alpha_0$, $\ldots$, $\alpha_{j-1}$ 
should be such that the state
is orthogonal to $T_{j-1}$ and, in particular, orthogonal 
to states $\ket{\psi^{i, 0}_{i_1, \ldots, i_l}}$ for 
$l\in\{0, \ldots, j-1\}$. By writing out
$\lbra \psi_0 \ket{\psi^{i, 0}_{i_1, \ldots, i_l}}=0$, we get
\begin{equation}
\label{eq-system} 
\sum_{m=l}^j \alpha_m {N-j-1 \choose K-m} {j-l \choose m-l} =0   .
\end{equation}
To show that, we first note that
$\ket{\psi^{i, 0}_{i_1, \ldots, i_l}}$ is a uniform superposition
of all $\ket{x}$, $|x|=K$, $x_i=0$, $x_{i_1}=\ldots=x_{i_l}=1$.
If we want to choose $x$ subject to those constraints
and also satisfying $|\{i_1, \ldots, i_j\}\cap \{t:x_t=1\}|=m$,
we have to set $x_t=1$ for $m-l$ different $t\in\{i_{l+1}, \ldots, i_j\}$
and for $K-m$ different $t\notin\{i, i_1, \ldots, i_j\}$. 
This can be done in ${j-l \choose m-l}$
and ${N-j-1 \choose K-m}$ different ways, respectively.

By solving the system of equations (\ref{eq-system}),
we get
that the only solution is
\begin{equation}
\label{eq-solution} 
 \alpha_m=(-1)^{j-m}
\frac{{N-j-1 \choose K-j}}{{N-j-1 \choose K-m}} \alpha_j 
.\end{equation}

Let $\ket{\psi'_0}=\frac{\ket{\psi_0}}{\|\psi_0\|}$ be the normalized
version of $\ket{\psi_0}$. Then,
\[ \ket{\tilde{\psi}^{i, 0}_{i_1, \ldots, i_j}} = 
\lbra \psi'_0\ket{\psi^{i, 0}_{i_1, \ldots, i_j}} \ket{\psi'_0} ,\]
\begin{equation}
\label{eq-2807} 
\| \tilde{\psi}^{i, 0}_{i_1, \ldots, i_j} \| = 
\lbra \psi'_0\ket{\psi^{i, 0}_{i_1, \ldots, i_j}} = 
\frac{\lbra \psi_0\ket{\psi^{i, 0}_{i_1, \ldots, i_j}}}{\|\psi_0\|} 
\end{equation}
First, we have 
\[ \lbra \psi_0 \ket{\psi^{i, 0}_{i_1, \ldots, i_j}} = 1 ,\]
because $\ket{\psi^{i, 0}_{i_1, \ldots, i_j}}$ consists of 
${N-j-1 \choose K-j}$
basis states $\ket{x}$, $x_i=0$, $x_{i_1}=\ldots=x_{i_j}=1$, each of which
has amplitude $1/\sqrt{{N-j-1 \choose K-j}}$
in both $\ket{\psi_0}$ and $\ket{\psi^{i, 0}_{i_1, \ldots, i_j}}$.
Second,
\[
\|\psi_0 \|^2 = \sum_{m=0}^j {j \choose m} {N-j-1 \choose K-m} \alpha_m^2 =
\sum_{m=0}^j {j \choose m} \frac{{N-j-1 \choose K-j}^2}{
{N-j-1 \choose K-m}} \alpha_j^2 \]
\[
=
\sum_{m=0}^j {j \choose m} \frac{{N-j-1 \choose K-j}}{{N-j-1 \choose K-m}} =
\sum_{m=0}^j {j \choose m} \frac{(K-m)!(N-K+m-j-1)!}{(K-j)!(N-K-1)!} \]
\begin{equation}
\label{eq-2807a} 
= \sum_{m=0}^j {j \choose m} 
\frac{(K-m) \ldots (K-j+1)}{(N-K-1) \ldots (N-K+m-j)}
\end{equation}
with the first equality following because there are 
${j \choose m}{N-j-1\choose K-m}$ vectors $x$ such that
$|x|=K$, $x_i=0$, $x_t=1$ for $m$ different $t\in\{i_1, \ldots, i_j\}$
and $K-m$ different $t\notin\{i, i_1, \ldots, i_j\}$,
the second equality following from equation 
(\ref{eq-solution}) and the third equality following from our choice 
$\alpha_j=1/\sqrt{{N-j-1 \choose K-j}}$. 

We can similarly calculate $\| \tilde{\psi}^{i, 1}_{i_1, \ldots, i_j} \|$.
We omit the details and just state the result. The counterpart of
equation (\ref{eq-2807}) is
\[  \| \tilde{\psi}^{i, 1}_{i_1, \ldots, i_j} \| = 
\frac{\lbra \psi_1\ket{\psi^{i, 1}_{i_1, \ldots, i_j}}}{\|\psi_1\|} ,\]
with $\ket{\psi_1}$ being the counterpart of $\ket{\psi_0}$:
\[ \ket{\psi_1}=\sum_{m=0}^j \alpha_m \mathop{\sum_{x:|x|=K, x_i=1}}_{
|\{i_1, \ldots, i_j\}\cap \{l:x_l=1\}| = m} \ket{x} ,\]
with $\alpha_0=1/\sqrt{{N-j-1 \choose K-j-1}}$.
Similarly as before, we get 
$\lbra \psi_1\ket{\psi^{i, 1}_{i_1, \ldots, i_j}}=1$
and 
\[ \|\psi_1 \|^2 = 
\sum_{m=0}^j {j \choose m} 
\frac{{N-j-1 \choose K-j-1}}{{N-j-1 \choose K-m-1}} 
\]
\begin{equation}
\label{eq-2807b} 
=\sum_{m=0}^j {j \choose m} 
\frac{(K-m-1) \ldots (K-j)}{(N-K) \ldots (N-K+m-j+1)}
\end{equation}
Each term in (\ref{eq-2807a}) is 
$\frac{(K-m)(N-K+m-j)}{(K-j)(N-K)}$
times the corresponding term in equation (\ref{eq-2807b}). 
We have
\[ \frac{K-m}{K-j} \frac{N-K+m-j}{N-K} \leq  \frac{K}{K/2} \cdot 2 = 4, \]
because $j\leq K/2$ and $N-K+m-j\leq N-K$ (because of $m\leq j$).
Therefore, $\|\psi_0\|^2 \leq 4 \|\psi_1\|^2$
which implies 
\[ \| \tilde{\psi}^{i, 0}_{i_1, \ldots, i_j} \| = 
\frac{1}{\|\psi_0\|}\geq \frac{1}{\sqrt{4} \|\psi_1\|} = 
\frac{1}{2} \| \tilde{\psi}^{i, 1}_{i_1, \ldots, i_j} \| .\]
\qed

{\bf Acknowledgment.}
I would like to thank Robert \v Spalek and Ronald de Wolf for very helpful comments
on a draft of this paper.

\end{document}